\documentclass[12pt,aps,prd,preprint,tightenlines,superscriptaddress,
   showpacs,nofootinbib]{revtex4}
\newcommand{\PRE}[1]{{#1}} 

\usepackage{bm}
\usepackage{graphicx}

\graphicspath{{figs/}}

\newcommand{\sigmaSI}{\sigma_{\rm SI}}
\newcommand{\sigmaSD}{\sigma_{\rm SD}}

\newcommand{\ifb}{\text{fb}^{-1}}

\newcommand{\gev}{\text{GeV}}
\newcommand{\tev}{\text{TeV}}
\newcommand{\pb}{\text{pb}}

\newcommand{\m}{\text{m}}
\newcommand{\km}{\text{km}}
\newcommand{\kg}{\text{kg}}

\newcommand{\eqref}[1]{Eq.~(\ref{#1})}

\newcommand{\Dsl}[1]{\slash\hskip -0.20 cm #1}

\newcommand{\be}{\begin{equation}}
\newcommand{\ee}{\end{equation}}
\newcommand{\bea}{\begin{eqnarray}}
\newcommand{\eea}{\end{eqnarray}}
\newcommand{\lsim}{\lower.7ex\hbox{$\;\stackrel{\textstyle<}{\sim}\;$}}
\newcommand{\gsim}{\lower.7ex\hbox{$\;\stackrel{\textstyle>}{\sim}\;$}}

\begin{document}

\preprint{UH-511-1166-2011}
\preprint{UTTG-06-11}
\preprint{TCC-008-11}

\title{
\PRE{\vspace*{1.3in}}
Detection Prospects for Majorana Fermion WIMPless Dark Matter
\PRE{\vspace*{0.3in}}
}

\author{Keita Fukushima}
\affiliation{Department of Physics and Astronomy, University of
Hawai'i, Honolulu, HI 96822, USA
\PRE{\vspace*{.1in}}
}

\author{Jason Kumar}
\affiliation{Department of Physics and Astronomy, University of
Hawai'i, Honolulu, HI 96822, USA
\PRE{\vspace*{.1in}}
}

\author{Pearl Sandick%
\PRE{\vspace*{.4in}}
}
\affiliation{Theory Group and Texas Cosmology Center, The University of Texas at Austin, TX 78712, USA
\PRE{\vspace*{.5in}}
}

\date{March 2011}

\begin{abstract}
\PRE{\vspace*{.3in}} We consider both velocity-dependent and velocity-independent contributions
to spin-dependent (SD) and spin-independent (SI) nuclear scattering (including one-loop corrections)
of WIMPless dark matter, in the case where
the dark matter candidate is a Majorana fermion.  We find that spin-independent scattering arises
only from the mixing of exotic squarks, or from velocity-dependent terms.  Nevertheless (and contrary to
the case of MSSM neutralino WIMPs), we find a class of models which cannot be detected through SI
scattering, but can be detected at IceCube/DeepCore through SD scattering.  We
study the detection prospects for both SI and SD detection strategies for a large range of
Majorana fermion WIMPless model parameters.
\end{abstract}

\maketitle

\section{Introduction}

WIMPless dark matter~\cite{Feng:2008ya,Feng:2009mn}
is a versatile scenario in which the dark matter candidate is a
hidden sector particle whose mass is at the hidden sector soft SUSY-breaking
scale.  The central feature of this scenario is that the dark matter
candidate is a thermal relic which naturally has approximately the correct relic density to
explain cosmological evidence of dark matter.
This is a very robust result which is essentially determined by
dimensional analysis, and does not depend on the details
of the hidden sector.  WIMPless dark matter models thus provide a wealth
of detection signatures which differ significantly from typical WIMP models,
spanning a wide range of dark matter masses (for either a bosonic or fermionic
dark matter candidate).

WIMPless dark matter couples to Standard Model matter multiplets through Yukawa
couplings to a 4th generation multiplet, yielding signals observable by
direct detection experiments, through indirect detection, and at colliders.
A variety of interesting signatures of WIMPless dark matter have been studied
in the case where the candidate is a scalar~\cite{Feng:2008dz,Feng:2008qn,McKeen:2009rm,Alwall:2010jc,Zhu:2011dz,Feng:2011vu},
with a particular focus on the limit of low mass and large spin-independent scattering
cross-section ($\sigmaSI$), i.e., the region of parameter-space nominally preferred
by the data from the DAMA, CoGeNT and CRESST experiments~\cite{lowmassdata}.  These
include signatures from direct detection experiments~\cite{Feng:2008dz,McKeen:2009rm,Feng:2011vu},
hadron colliders~\cite{Alwall:2010jc}, neutrino detectors~\cite{Feng:2008qn},
gamma ray measurements~\cite{Feng:2008dz},
measurements of invisible $\Upsilon$-decays at B-factories and new contributions to $b-s$ mixing~\cite{McKeen:2009rm}.

But there are other interesting examples of WIMPless dark matter that provide very
different detection signatures which will soon be probed by experiments.  One of the
most interesting cases is when the WIMPless dark matter candidate is a Majorana fermion.
In this case, the
tree-level scattering cross-section can have both spin-dependent ($\sigmaSD$) and
spin-independent ($\sigmaSI$) components.  While it is possible for scattering to be entirely
spin-dependent, it cannot be entirely spin-independent.

Dark matter detection experiments can be sensitive to
$\sigmaSI$ and/or $\sigmaSD$.  The best experimental sensitivity to both types of
scattering is for dark matter mass $m_X \sim {\cal O}(100)~\gev$.  For this mass scale,
direct detection experiments like SuperCDMS, XENON 100/1T and LUX are expected to provide the best
sensitivity to $\sigmaSI$, while neutrino experiments like IceCube/DeepCore are
expected to have the best sensitivity to $\sigmaSD$\footnote{The sensitivity of neutrino detectors to
dark matter annihilation products depends, of course, on the branching fractions to different Standard Model final states.}.
Typically, direct detection experiments
are much more sensitive to $\sigmaSI$, due to the $A^2$ enhancement
arising from coherent nuclear scattering.  As a result, current
bounds on $\sigmaSI$ are roughly 4 orders of magnitude tighter than those
for $\sigmaSD$, and future experimental sensitivity to $\sigmaSI$ is expected to far exceed
sensitivity to $\sigmaSD$.  Indeed, experimental sensitivity to even velocity-suppressed
spin-independent couplings can rival sensitivity to spin-dependent couplings~\cite{Freytsis:2010ne}.
For many dark matter models, this implies that $\sigmaSI$ is
most relevant for detection.

For the MSSM, a scan of parameters (without the assumption of gaugino mass
unification) has found models for which $\sigmaSD$ can potentially be measured, but
for which velocity-independent contributions to $\sigmaSI$ are too
small to be detected with current or future experiments~\cite{Moulin:2005sx}.
However, such models tend to be focused at low-mass; the scan in~\cite{Moulin:2005sx} did
not find any such models with $m_X \gsim 200~\gev$.  Since the couplings of the
MSSM dictate the relative contribution of Higgs, Z, squark and axion exchange to
neutralino-nucleon scattering, it is difficult to entirely decouple spin-dependent
couplings from spin-independent couplings.  Moveover, MSSM models that exhibit spin-dependent
scattering will also exhibit velocity-dependent spin-independent scattering; models which
can be detected through spin-dependent scattering at current direct detection experiments can
also be detected through even velocity-suppressed spin-independent scattering at direct
detection experiments operating now or in the near future~\cite{Freytsis:2010ne}.

But for WIMPless dark matter, potential dark matter-nucleon interactions are
more limited (exchange of a 4th generation multiplet).
For Majorana fermion WIMPless models, the spin-dependent and spin-independent interactions
can be decoupled for a wide range of dark matter mass.  Since these models can have very small or vanishing
$\sigmaSI$, it is quite possible that for many such models, the key to detection
is spin-dependent scattering.  Detection prospects for Majorana fermion WIMPless
dark matter thus depend sensitively on the interplay between spin-dependent and spin-independent contributions
to dark matter-nucleus scattering.

In this paper, we consider the detection prospects for Majorana fermion WIMPless
dark matter through both spin-dependent and spin-independent scattering.  In section
II, we describe the interactions of Majorana fermion WIMPless dark matter, in
section III we review current and future experimental sensitivity to
dark matter-nucleon scattering, in section IV we describe detection prospects for
Majorana fermion WIMPless dark matter, and we conclude in section V.

\section{Interactions of Majorana Fermion WIMPless Dark Matter}
\label{sec:interactions}

The WIMPless dark matter candidate is a hidden sector particle ($X$)
at the hidden sector soft SUSY-breaking scale.
The main feature of these models is that $g_X^2 / m_X \sim g_{weak}^2 / m_{weak}$, where
$g_X$ is the hidden sector gauge coupling, and $m_X$ is the dark matter mass.
Since $\langle \sigma_{ann.} v \rangle \propto (g^4 / m^2)$, the WIMPless
dark matter candidate has roughly the same annihilation cross-section as a WIMP,
implying that the WIMPless candidate has approximately the correct relic density.
A simple implementation of this model arises in gauge-mediated SUSY-breaking\footnote{WIMPless models
can also be constructed in the context of anomaly-mediated supersymmetry-breaking~\cite{Feng:2011ik}.},
where both the hidden sector and MSSM sector receive the effects of SUSY-breaking
from a common SUSY-breaking sector.  Both
the MSSM and hidden sector soft scales are related to their gauge
couplings by the vevs of the same spurion field,
$\langle \Phi \rangle = M + \theta^2 F$, through the relation
\bea
{g_X^2 \over m_X} , {g_{weak}^2 \over m_{weak}} &\propto& {M \over F}.
\eea

The case where the dark matter candidate is a Majorana fermion has previously been
discussed in~\cite{Barger:2010ng}.  Majorana fermion WIMPless dark matter couples
to Standard Model quarks through the Yukawa couplings
\bea
V &=& \lambda_{Li} (\bar X P_L q_i)\tilde Y_L
+ \lambda_{Ri} (\bar X P_R q_i)\tilde Y_R +h.c.
\eea
where $q_i$ are MSSM quarks and $i$ is a flavor index.
The WIMPless dark matter candidate, $X$, is neutral under Standard Model symmetries; it is charged only under the
hidden sector symmetry that stabilizes it (for simplicity, this symmetry is
assumed to be discrete).
The $\tilde Y_{L,R}$  are exotic
scalar connector particles which are charged under both the MSSM and the hidden sector
symmetry.    Gauge-invariance thus implies that the
$\tilde Y_{1,2}$ are in a 4th generation quark multiplet.

The $\tilde Y_{L,R}$ need not be mass eigenstates.  In general, the mass eigenstates
$\tilde Y_{1,2}$ are related to $\tilde Y_{L,R}$, through the mixing angle, $\alpha$;
\bea
\tilde Y_L &=& \tilde Y_1 \cos \alpha + \tilde Y_2 \sin \alpha
\nonumber\\
\tilde Y_R &=& -\tilde Y_1 \sin \alpha + \tilde Y_2 \cos \alpha.
\eea

If the dark matter candidate is a Majorana fermion, then the above Yukawa couplings
permit scattering from Standard Model quarks via tree-level $s$- or $u$-channel
exchange of the 4th generation squarks, $\tilde Y_{L,R}$.  The dark matter candidate, $X$,
can also annihilate to Standard Model particles through exchange of a 4th generation multiplet.

In~\cite{Barger:2010ng}, it was assumed that the 4th generation squarks $\tilde Y_L$
and $\tilde Y_R$ did not mix, but this need not be true. Assuming $m_{\tilde Y_1} < m_{\tilde Y_2}$,
dark matter-nucleon scattering is mediated by the effective operator
\bea
\label{operatoreq}
{\cal O} & =&
 \alpha_{1i} (\bar X \gamma^\mu \gamma^5 X) (\bar q_i \gamma_\mu q_i) +
\alpha_{2i}
(\bar X \gamma^\mu \gamma^5 X)(\bar q_i \gamma_\mu \gamma^5 q_i)
+\alpha_{3i}(\bar X X)(\bar q_i q_i )
\nonumber\\
&\,&
+\alpha_{4i} (\bar X \gamma^5 X) (\bar q_i  \gamma^5 q_i )
+ \alpha_{5i} (\bar X X) (\bar q_i  \gamma^5 q_i)
+\alpha_{6i} (\bar X \gamma^5 X) (\bar q_i   q_i),
\label{eq:alphas}\eea
with
\bea
\alpha_{1i} &= &
\left[{|\lambda_{Li}^2| \over 8}
\left( {\cos^2 \alpha \over m_{\tilde Y_1}^2 -m_X^2}
+ {\sin^2 \alpha \over m_{\tilde Y_2}^2 -m_X^2} \right)
-{|\lambda_{Ri}^2| \over 8}
\left({\cos^2 \alpha \over m_{\tilde Y_2}^2 -m_X^2}
+ {\sin^2 \alpha \over m_{\tilde Y_1}^2 -m_X^2} \right) \right]
\nonumber\\
\alpha_{2i} &=& {|\lambda_{Li}^2| \over 8}
\left( {\cos^2 \alpha \over m_{\tilde Y_1}^2 -m_X^2}
+ {\sin^2 \alpha \over m_{\tilde Y_2}^2 -m_X^2} \right)
+{|\lambda_{Ri}^2| \over 8}
\left({\cos^2 \alpha \over m_{\tilde Y_2}^2 -m_X^2}
+ {\sin^2 \alpha \over m_{\tilde Y_1}^2 -m_X^2} \right)
\nonumber\\
\alpha_{3i,4i} &=& {Re(\lambda_{Li} \lambda_{Ri}^*) \over 4}(\cos \alpha \sin \alpha)
\left[{1\over m_{\tilde Y_1}^2 -m_X^2} -{1\over m_{\tilde Y_2}^2 -m_X^2} \right]
\nonumber\\
\alpha_{5i,6i} &=& {\imath Im(\lambda_{Li} \lambda_{Ri}^*) \over 4}(\cos \alpha \sin \alpha)
\left[{1\over m_{\tilde Y_1}^2 -m_X^2} -{1\over m_{\tilde Y_2}^2 -m_X^2} \right]
\eea
where we have used the notation of~\cite{DDnotation}.  As shown there,
$\alpha_2$ is the coefficient of a pseudovector coupling which mediates spin-dependent scattering,
while $\alpha_3$ is the coefficient of a scalar coupling which mediates spin-independent
scattering.  These are the only couplings which are not velocity-suppressed.  $\alpha_4$ and
$\alpha_5$ are the coefficients of spin-dependent couplings which are also velocity-suppressed and
will not be relevant for the detection prospects studied here.
However, $\alpha_1$ and $\alpha_6$ are the coefficients of spin-independent couplings which are
velocity-suppressed; though suppressed, the greater sensitivity of direct detection experiments to
spin-independent couplings may make them relevant for detection purposes.

The dark matter-nucleus scattering cross-sections can then be written as
\bea
\sigmaSI &=& \sigmaSI^{(1)} + \sigmaSI^{(3)} + \sigmaSI^{(6)}
\nonumber\\
\sigmaSD &=& \sigmaSD^{(2)},
\eea
with
\bea
\sigmaSI^{(1)} &=&
{4 m_r^2 \over \pi}
\left[Z(\sum_i \alpha_{1i} B^p_i) +(A-Z)(\sum_i \alpha_{1i} B^n_i)\right]^2 {v^2 \over 2}
\nonumber\\
\sigmaSI^{(3)} &=&
{4 m_r^2 \over \pi}
\left[Z(\sum_i \alpha_{3i} B^p_i) +(A-Z)(\sum_i \alpha_{3i} B^n_i)\right]^2
\nonumber\\
\sigmaSI^{(6)} &=&
{4 m_r^2 \over \pi}
\left[Z(\sum_i \alpha_{6i} B^p_i) +(A-Z)(\sum_i \alpha_{6i} B^n_i)\right]^2 {m_r^2 v^2 \over 2m_X^2}
\nonumber\\
\sigmaSD^{(2)} &=& {16 m_r^2  \over \pi }
\left[\sum_i \alpha_{2i} (\Delta_i^{(p)} \langle S_p \rangle
+\Delta_i^{(n)} \langle S_n \rangle ) \right]^2  {J+1 \over J}.
\eea
Note that $\sigmaSI^{(6)}$ is suppressed by an additional factor $({m_r / m_X})^2$, which
is always less than 1.
The integrated nuclear form factors are
\bea
B^p_u = B^n_d \simeq 12 \qquad B^p_d = B^n_u \simeq 6 \qquad B^{p,n}_s \simeq 4
\eea
and we take the spin form factors to be~\cite{spinfactors}
\bea
\Delta_u^{(p)} = \Delta_d^{(n)} = 0.84
\qquad
\Delta_d^{(p)} = \Delta_u^{(n)} = -0.43
\qquad
\Delta_s^{(p,n)} = -0.09\,.
\eea
If dark matter couples largely to heavy quarks, then its scattering is
largely spin-independent.
Henceforth, we will focus on the case where the dark matter couples largely
to the light quarks.  We assume that the dominant couplings are to first generation
quarks, and for simplicity, we will also assume $\lambda_{Ru} = \lambda_{Rd}=\lambda_R$.

The ratio of spin-independent to spin-dependent couplings is
bounded from above by ${\alpha_3 \over \alpha_2} \leq {1\over 2}$, where the
inequality is saturated in limit of real couplings with $\lambda_L = \lambda_R$,
maximal mixing ($\alpha = {\pi \over 4}$) and $m_{\tilde Y_2} \rightarrow \infty$. This maximum value of $\sigma_{SI}/\sigma_{SD}$ is obtained only outside the range of $\sigmaSI$ and $\sigmaSD$ considered here.

The analysis of~\cite{Barger:2010ng} corresponds to the case
where $\alpha_{1,3,6}=0$.  The dominant spin-independent coupling arises
from $\alpha_3$, and there are four limits in which $\alpha_3 \rightarrow 0$ :
\begin{itemize}
\item{$\lambda_L=0$ (or $\lambda_R=0$)}
\item{$\alpha =0$}
\item{$\m_{\tilde Y_1} = m_{\tilde Y_2}$}
\item{maximal CP-violation ($\arg(\lambda_L \lambda_R^*) =\pm {\pi \over 2}$)}
\end{itemize}
For the remainder of this work, we will focus on the case of real Yukawa couplings,
i.e.~no CP-violation.
In these limits, spin-independent scattering is necessarily velocity-suppressed.
But even these velocity-dependent terms can be made arbitrarily small.  For example,
in the limit
\be
\alpha \rightarrow 0 \,\,\, \textrm{and} \,\,\, {|\lambda_L^2| \over (m_{\tilde Y_1}^2-m_X^2)} = {|\lambda_R^2| \over (m_{\tilde Y_2}^2-m_X^2)}
\ee
one would have $\alpha_{1,3,6} \rightarrow 0$.  In this limit,
scattering detection prospects must rely entirely on detectors sensitive
to $\sigmaSD$.  As one deviates from these limits, detectors sensitive to $\sigmaSI$ can become relevant.

\subsection{One-loop corrections}

One should also consider one-loop scattering diagrams which could potentially
generate spin-independent scattering.  However, we shall find that in the case of
Majorana fermion WIMPless dark matter these are not relevant to direct detection
prospects.

We consider the limit where $\alpha \rightarrow 0$, since terms involving squark-mixing
will yield spin-independent scattering even at tree-level.  The only relevant one-loop
scattering diagrams we can write involve $s$- or $u$-channel exchange of a squark, along with
exchange of a photon or $Z$ between the Standard Model fermion lines.  Since we are considering
models that couple to 1st generation quarks, the $h$, $H$ and $A$ exchange diagrams are highly
suppressed.  Assuming the squark exchanged is $\tilde Y_L$, the matrix element for the relevant
one-loop diagrams is of the form
\bea
{\cal M} &\propto& g^2 \lambda_L^2  \int {d^4 p \over (2\pi)^4}
\langle Xf|[\bar X P_L (\Dsl p +m_f) \gamma^\mu f]
[\bar f \gamma_\mu (\Dsl p -m_f) P_R X]|Xf \rangle f_0(p),
\eea
where the factor $f_0(p)$ contains the momentum dependence
of the propagators in the loop diagram.  Note that the terms
proportional to $g^2 m_f^2$ will be heavily suppressed compared to the tree-level terms, and can be
dropped.
Furthermore, the matrix element will be of this form regardless of whether a photon or $Z$ is
exchanged; an additional $\gamma^5$ factor at one or more interaction vertices will not change
the matrix element except by an overall sign, since $\gamma^5 P_{L,R} = \pm P_{L,R}$.
We then find
\bea
{\cal M} &\propto&
g^2 \lambda_L^2 \langle Xf| (\bar X P_L \gamma^\nu \gamma^\mu f)
(\bar f \gamma_\mu \gamma_\nu P_R X) |Xf \rangle
\nonumber\\
&=& g^2 \lambda_L^2 \langle Xf|(\bar X P_L  f)(\bar f  P_R X) |Xf \rangle-
2g^2 \lambda_L^2 \langle Xf|(\bar X  \sigma^{\mu \nu} P_L f)
(\bar f \sigma_{\mu \nu} P_R X)|Xf \rangle.
\eea
After a Fierz transformation, the second term in brackets can be shown to vanish.
The first term in brackets is a pseudovector coupling,
and therefore corresponds to a loop-correction to $\sigmaSD$, which is
irrelevant for our purposes.  We thus see that there are no relevant one-loop contributions
to $\sigmaSI$, and spin-independent scattering arises only from squark-mixing and velocity-suppressed
contributions.

\section{Current and Prospective Bounds on $\sigma_{SI}$ and $\sigma_{SD}$}
\label{sec:bounds}

We now briefly review some of the detectors which are relevant to current and future bounds
on (sensitivities to) $\sigmaSI$ and $\sigmaSD$.  These detection bounds arise from direct
detection experiments, neutrino experiments, and also from colliders.  Direct detection experiments measure the energy of a nucleus recoiling
from a dark matter interaction, whereas neutrino experiments search for
the neutrino flux produced when two dark matter particles annihilate.  Direct detection experiments are sensitive to
both $\sigmaSI$ and $\sigmaSD$.  Although neutrino experiments are also sensitive to $\sigmaSI$, their
sensitivity does not rival that of direct detection experiments, whose sensitivity to $\sigmaSI$ is enhanced
by coherent scattering of the heavy nuclei of the detector.  But the sensitivity of neutrino experiments to
$\sigmaSD$ is very competitive with that of direct detection experiments.

The current leading bounds on $\sigmaSI$ at $m_X = 100~\gev$ are set by CDMS-II
(Cryogenic Dark Matter Search II)~\cite{Ahmed:2009zw} and XENON100~\cite{Aprile:2010um}.
CDMS-II utilizes germanium ($\sim 4.4~\kg$) and silicon ($\sim 1.1~\kg$) detectors, and measures
both ionization and phonon energy to distinguish signal events from background.
XENON100 utilizes liquid xenon ($62~\kg$ fiducial mass), and similarly measures both ionization and
scintillation yield to distinguish signal from background.  Current bounds from XENON100 have been set
using 11 days of data.  At the time of publication, XENON100 is collecting data,
and an upgrade to $\sim 1000~\kg$ fiducial mass (XENON1T)~\cite{Aprile:2002ef} is expected to begin operating around 2014.
Meanwhile, SuperCDMS is expected to begin operation in 2013-2015, using a germanium detector with  a target mass of
100 kg~\cite{Schnee:2005pj}.

GeODM 1.5T (Germanium Observatory for Dark Matter) will detect ionization and phonons in a $\sim 1500~\kg$
fiducial mass germanium target~\cite{Brink:2010uc}, and is expected to begin operating in 2017-2021.

The LUX (Large Underground Xenon) experiment is also a liquid xenon-based detector, with a 350 kg fiducial mass~\cite{Fiorucci:2009ak}.
This detector is
expected to begin operation in 2011.
In conjunction with ZEPLIN (Zoned Proportional scintillation in LIquid Noble gases),
upgrades to a fiducial mass of 3000 kg (LUX-ZEP 3T) and, finally,
20000 kg (LUX-ZEP 20T), are planned for the next several years~\cite{Fiorucci:2009ak}.

The DEAP/CLEAN family of detectors are purely liquid-phase noble gas scintillation detectors utilizing either neon or argon.
MiniCLEAN (Cryogenic Low Energy Astrophysics with Noble gases) runs with either $\sim 100$ kg fiducial
mass of liquid argon or $\sim 85$ kg fiducial mass of liquid neon as the target~\cite{McKinsey:2007zz}.
It is expected to be installed at SNOLAB this year.
DEAP-3600 (Dark matter Experiment in Argon using Pulse-shape discrimination) is a liquid argon detector with $\sim 1000$ kg fiducial mass,
also expected to be installed at SNOLAB this year~\cite{Boulay:2009zw}.

LUX-ZEP 20T would have the greatest sensitivity to $\sigmaSI$ obtainable from a zero-background
direct detection experiment ($\sim 10^{-12}~\pb$).
For smaller cross-sections, neutrino-nucleus scattering
becomes a significant and irreducible background~\cite{nuscatBG}.

Significant bounds on spin-dependent scattering are obtained from both direct detection experiments and from neutrino detectors
searching for dark matter annihilation in the sun.
The best current bound on $\sigmaSD^p$ from a direct detection experiment at $m_X = 100~\gev$ is from SIMPLE (Superheated Instrument for
Massive ParticLe Experiments)~\cite{simple}, which consists of superheated droplet detectors made of ${\rm C_2ClF_5}$ droplets in a gel matrix.  COUPP
(Chicagoland Observatory for Underground Particle Physics)~\cite{Behnke:2010xt}, which uses a bubble chamber filled with 3.5kg of ${\rm CF_3 I}$ to search
for dark matter nuclear recoils while rejecting electron recoil events, is also competitive.
The best current bound on $\sigmaSD^n$ from a direct detection experiment at $m_X=100$ GeV comes from XENON10~\cite{xenon10}, the precursor to XENON100.

Improved bounds on $\sigmaSD^p$ are possible with DMTPC (Dark Matter Time Projection Chamber), a ${\rm CF_4}$ gas scintillator.
DMTPC currently operates with a 10 L fiducial volume~\cite{Battat:2010ip}, but
it is anticipated that DMTPC will be upgraded to a $1~\m^3$ fiducial volume (DMTPC-ino), with a further possible upgrade
to Large DMTPC with $10^2-10^3~\kg$ fiducial mass~\cite{Sciolla:2009fb,Battat:2010ip}.

Super-Kamiokande is a 22.5 kT water Chernekov detector and can detect charged-current interactions of neutrinos
arising from dark matter annihilation in the Sun~\cite{Desai:2004pq}.  It has accumulate over
3000 live-days of data already.
Similarly, IceCube uses $\sim 1~\km^3$ of deep and ultra-transparent Antarctic ice as a Cherenkov detector
of neutrinos~\cite{Braun:2009fr}.
The DeepCore extension to IceCube includes a denser array of digital
optical modules, whose installation was completed in January 2010.
This denser array allows IceCube/DeepCore to be
sensitive to lower-mass dark matter ($m_X > 35~\gev$).  One should note that the sensitivity of neutrino detectors
to  $\sigmaSD$ is somewhat model-dependent, and in particular depends on the dark matter annihilation channel.
The Super-Kamiokande bound assumes dark matter annihilation to $b$-quarks.
The prospective IceCube/DeepCore sensitivity cited here is the ``hard" channel reported by IceCube/DeepCore, which is the assumption of annihilation
entirely to $\tau$ leptons for $m_X<80~\gev$, and annihilation to $W$ bosons for $m_X > 80~\gev$ (although the bounds
found in~\cite{Barger:2010ng} are tighter than those reported in~\cite{Braun:2009fr}), for 1800 live-days of data.
The prospective bounds from these neutrino detectors are only relevant here if the dark matter candidate has a significant
annihilation branching fraction to heavy particles.
This is certainly reasonable, as the WIMPless candidate is a Majorana fermion, so the
cross section for annihilation to light fermions is chirality suppressed.

Collider experiments, such as the Tevatron and Large Hadron Collider (LHC) can also constrain the dark matter-nucleon scattering
cross section~\cite{Feng:2005gj,Goodman:2010yf}\footnote{Note that the analysis of~\cite{Goodman:2010yf} assumes that the pseudovector coupling
of dark matter is flavor-independent, whereas we are considering WIMPless models which couple only to first generation quarks.  We expect
that the coupling of dark matter to second and third generation quarks has a negligible effect on collider sensitivity to the pseudovector coupling.}.
The effective operators in eq.~\ref{operatoreq} allow two quarks to annihilate to dark matter, which contributes to the
process $p\bar p (pp) \rightarrow XX+ jets$, where the jets arise from initial state radiation.
Collider searches for this process can thus constrain the operator coefficients $\alpha_i$.

CDF has performed monojet searches with $1~\ifb$ of data~\cite{Aaltonen:2008hh}.   The search looks for events with
missing $p_T > 80~\gev$, a leading jet with $p_T > 80~\gev$, and the constraints that a second jet (if present) must
have $p_T < 30~\gev$ and that there be no other jets with $p_T > 20~\gev$.
This search bounds the new physics contribution to this signature
at $\sigma_{NP} < 0.664~\pb$ at the $2\sigma$-confidence level.
If one assumes that only $\alpha_3$ is non-zero, the CDF bound on $\alpha_3$ corresponds to a bound
on $\sigmaSD$ (for $m_X \sim 100~\gev$, similar constraints on the effective
operators contributing to spin-independent scattering are
not competitive with bounds from direct detection experiments).  Tevatron bounds on this pseudovector exchange effective
operator are quite stringent~\cite{Goodman:2010yf}.

In~\cite{Goodman:2010yf}, the ability of the LHC to probe the pseudovector exchange scattering operator
with $100~\ifb$ of data was also studied.  This studied assumed a signal of jets and missing $p_T > 500~\gev$, where
the background was determined from~\cite{Vacavant:2001sd}.
The sensitivity of the LHC to this operator will be very competitive with
that from direct detection and neutrino experiments.

However, as with neutrino experiments, the bounds from collider experiments are also somewhat model-dependent.  In
particular, these bounds assume the validity of the effective operator analysis at collider energies.  These bounds
are strictly valid in the limit where the exotic squark masses $m_{\tilde Y{1,2}}$ are much larger energy of the hard
scattering process; if the squark masses are comparable to collider energies, these bounds can be weakened.

\section{Detection Prospects}

In this section, we present the
results of a WIMPless dark matter parameter space scan, followed by an analysis of the three limits described in
Section~\ref{sec:interactions} in which $\sigmaSI \rightarrow 0$ (neglecting the limit of maximal CP-violation).

Figure~\ref{fig:bigscan} shows $\sigmaSI$ and $\sigmaSD$ for a scan of the parameter space of Majorana fermion
WIMPless dark matter, where the parameters scanned over are
\bea
&{0 \leq \alpha \leq \pi / 4}\nonumber\\
&{0 \leq \lambda_{L,R} \leq \sqrt{4\pi}}\nonumber\\
&{300\,\textrm{GeV} < m_{\tilde Y_1} < 2 \,\textrm{TeV}}\nonumber\\
&{m_{\tilde Y_1} < m_{\tilde Y_2} < 2 \,\textrm{TeV}}
\eea
and we take $m_X=100$ GeV. The scan includes $2 \times 10^5$ model points, though a significant fraction lie at
larger $\sigmaSI$ than is shown in Fig.~\ref{fig:bigscan}.
For the velocity-dependent terms in the cross-section, we approximate $v = 220$ km/s.  In a more
detailed calculation, one would convolve the cross-section against a velocity distribution to determine an
event rate at any given experiment, which in turn would determine the sensitivity at that experiment.  Our
approximation is sufficient for the purpose here, and allows us to determine the sensitivity of any
given experiment to Majorana fermion WIMPless dark matter using publicly available bounds.

\begin{figure}[tb]
\begin{center}
\includegraphics*[width=.95\textwidth]{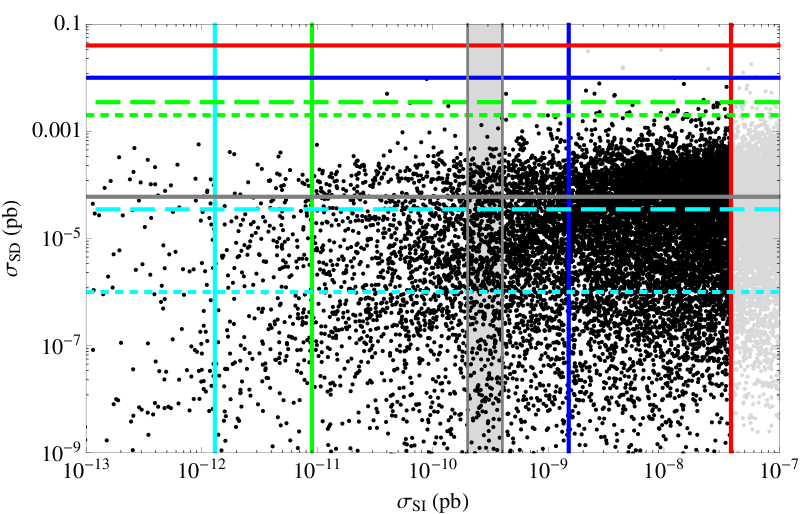}
\end{center}
\vspace*{-.25in}
\caption{Possible values of $\sigmaSI$ and $\sigmaSD$ for $0 \leq
\lambda_L, \lambda_R \leq \sqrt{4\pi}$, $0 \leq \alpha \leq {\pi \over 4}$
and 300 GeV $< m_{\tilde Y_1} < 2$ TeV, and $m_{\tilde Y_1} < m_{\tilde Y_2} < 2$ TeV ($m_X = 100~\gev$).
The dark (black) points represent viable models, the light (grey) points represent models that are ruled out by direct
detection experiments.
Also shown are the current limits and projected sensitivities of
direct detection experiments for $\sigmaSI$ (vertical lines) and $\sigmaSD$ (horizontal lines). Constraints on $\sigmaSI$ are (listed from
right to left): the current limit from CDMS II~\cite{Ahmed:2009zw} and XENON100~\cite{Aprile:2010um} (red solid);
mini-CLEAN~\cite{DMplotter} (blue solid); the approximate sensitivity of SuperCDMS 100kg at SNOLAB~\cite{Schnee:2005pj}, DEAP-3600~\cite{DMplotter},
XENON100 upgraded~\cite{Aprile:2010um}, and LUX with 300 days of exposure~\cite{lux300} (grey band); XENON1T~\cite{Aprile:2002ef} with 3 years of exposure (green solid), and
LUX/ZEP 20T~\cite{DMplotter} (cyan solid).
Constraints on $\sigmaSD$ from direct detection experiments are shown as the solid lines (listed from top to bottom): the current limit from SIMPLE~\cite{simple}
(red solid) for scattering on protons; the current limit from XENON10~\cite{xenon10} (blue solid) for scattering on neutrons, and the projected sensitivity of
Large DMTPC~\cite{Sciolla:2009fb} (grey solid).  Indirect constraints on $\sigmaSD$ from neutrino experiments are shown as dashed lines (listed from top to bottom):
Super-Kamiokande~\cite{Desai:2004pq} (green dashed); and the projected sensitivity of IceCube/DeepCore~\cite{Braun:2009fr} with 1800 days of data
(cyan dashed).  The sensitivity to $\sigmaSD$ of collider experiments, assuming contact interactions, is shown as dotted lines (listed from top to bottom): the
Tevatron~\cite{Goodman:2010yf} with $1~\ifb$ of data (green dotted); and the projected sensitivity of the LHC~\cite{Goodman:2010yf} with $100~\ifb$ of data (cyan dotted).
\label{fig:bigscan}}
\end{figure}

We note that the priors for this scan are linear in $\alpha$, $\lambda_{L,R}$ and $m_{\tilde Y_{1,2}}$.
The current limits and projected sensitivities of direct detection experiments
for $\sigmaSI$ are shown as vertical lines, and those for $\sigmaSD$ are shown as horizontal lines.
The region of the plane above the horizontal solid blue line is currently excluded by non-observation of spin-dependent scattering on neutrons by XENON10
(the limit on $\sigmaSD^p$ from SIMPLE is shown in red),
while the region of the plane to the right of the vertical solid red line is excluded by non-observation of spin-independent
scattering by CDMS-II and XENON10.
Spin-(in)dependent cross sections above (to the right of) a particular sensitivity will be probed by the corresponding experiment.

Because of the greater sensitivity to $\sigmaSI$ of
detectors such as SuperCDMS, XENON1T and LUX, most models in this scan either have already been excluded, or
can be detected through spin-independent scattering by future searches.
The dominant contribution to spin-independent scattering is the velocity-independent term $\sigmaSI^{(3)}$.  The velocity-dependent terms $\sigmaSI^{(1,6)}$ are
suppressed by $\sim 6$ orders of magnitude, and only become relevant to detection
prospects in regions of parameter space where $\sigmaSI^{(3)} \sim 0$.
Nevertheless, there are some
models which cannot be detected through $\sigmaSI$ with currently planned
detectors, but can be detected through $\sigmaSD$ at IceCube/DeepCore, or at the LHC\footnote{Sensitivities of neutrino and collider experiments to $\sigmaSD$ are shown as dashed and dotted lines, respectively, in Figs.~\ref{fig:bigscan}-\ref{fig:velocityonly}.}. These models lie above the cyan dashed IceCube/DeepCore
sensitivity line (or the cyan dotted LHC sensitivity line),
and at very low $\sigmaSI$.
Note, however, that the collider bounds from the Tevatron and the LHC (dotted lines) are only strictly valid in the limit where the
squark mass is much larger than the collider energy scale.  This will be true for models $m_{\tilde Y_{1,2}} \sim 2~\tev$, but
some of these collider bounds can be substantially weakened if the squarks are lighter.


\begin{figure}[ht]
\begin{center}
\mbox{\includegraphics[width=0.48\textwidth]{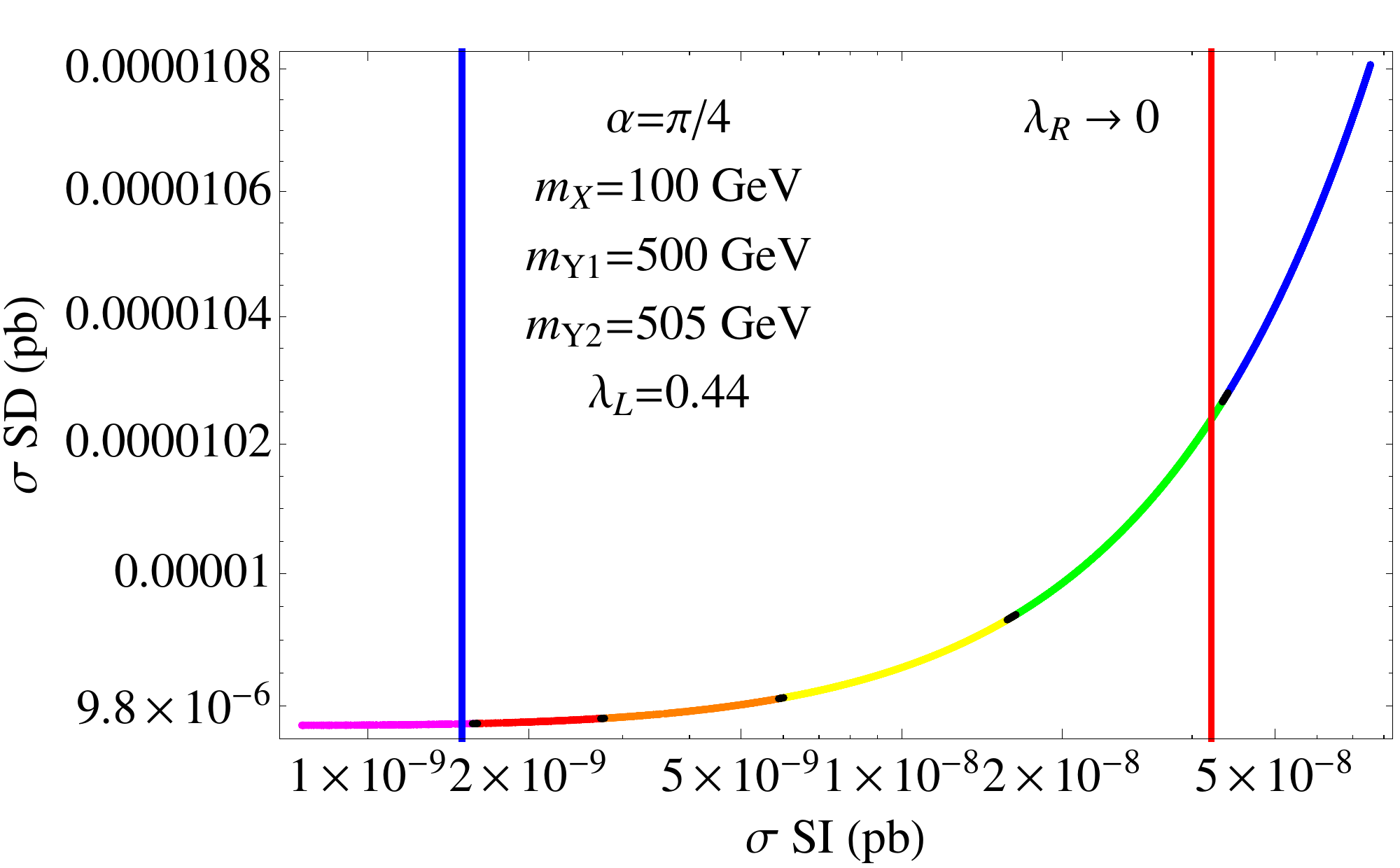}}\hspace{2mm}
\mbox{\includegraphics[width=0.48\textwidth]{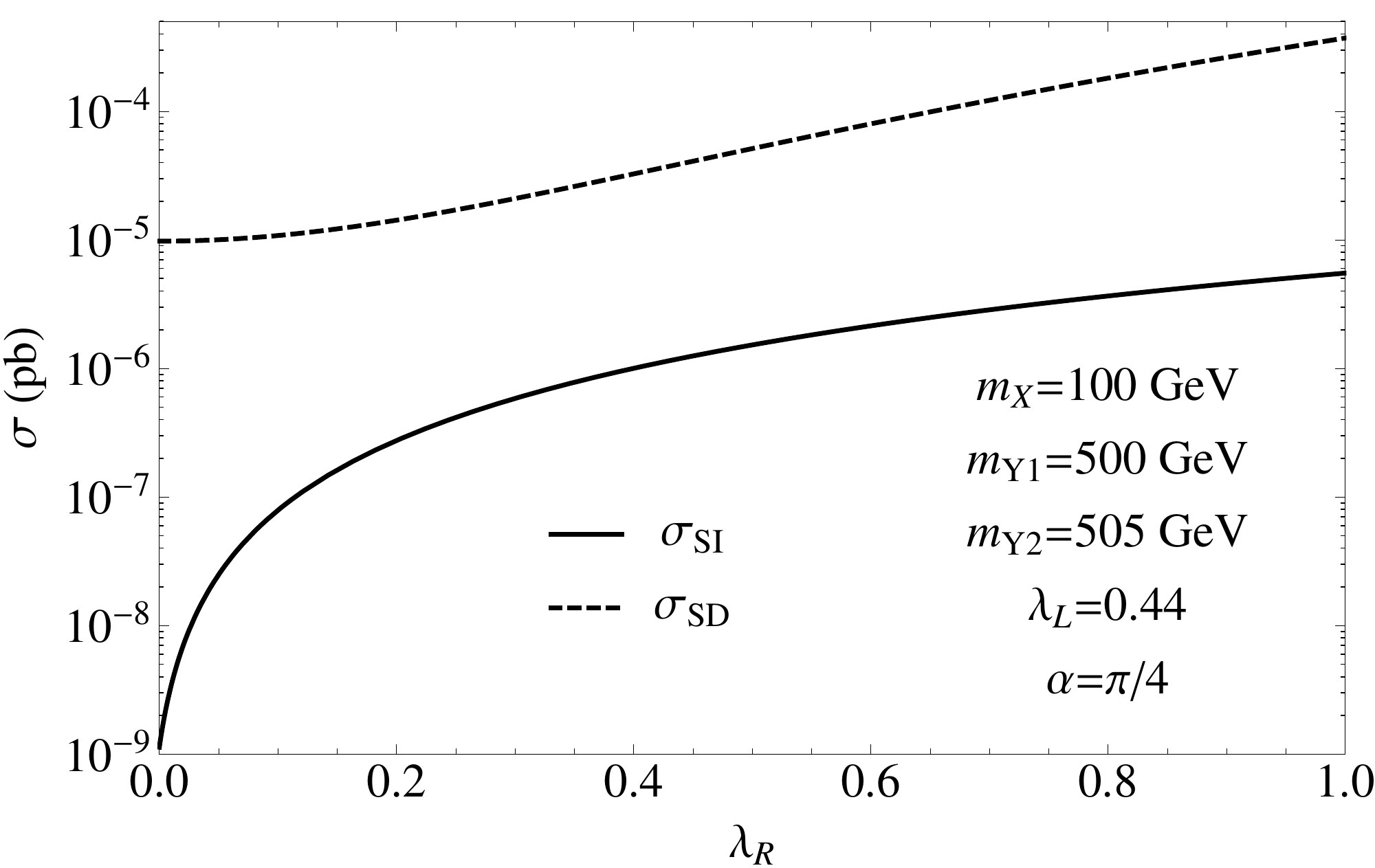}}\vspace{4mm}
\mbox{\includegraphics[width=0.48\textwidth]{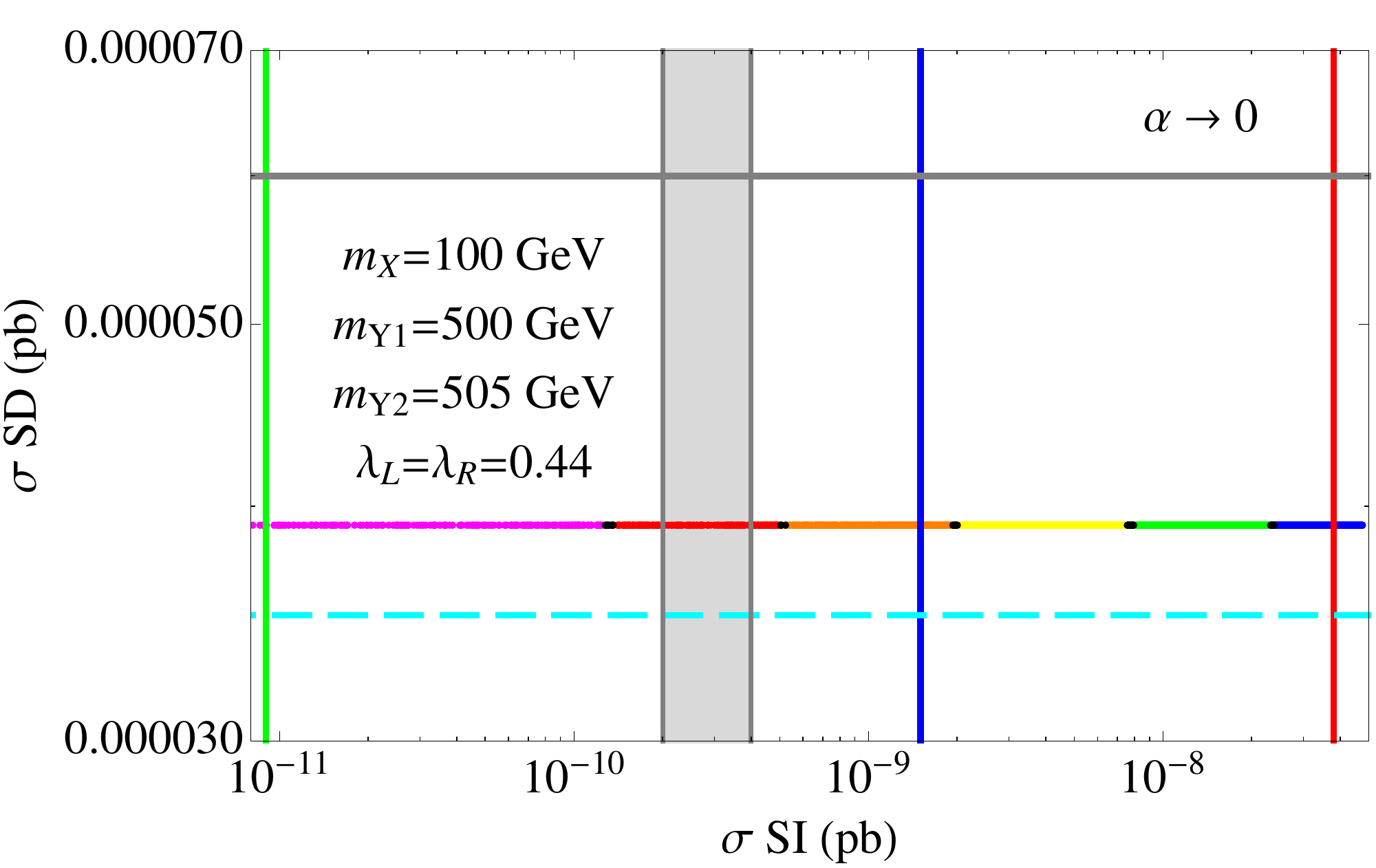}}\hspace{2mm}
\mbox{\includegraphics[width=0.48\textwidth]{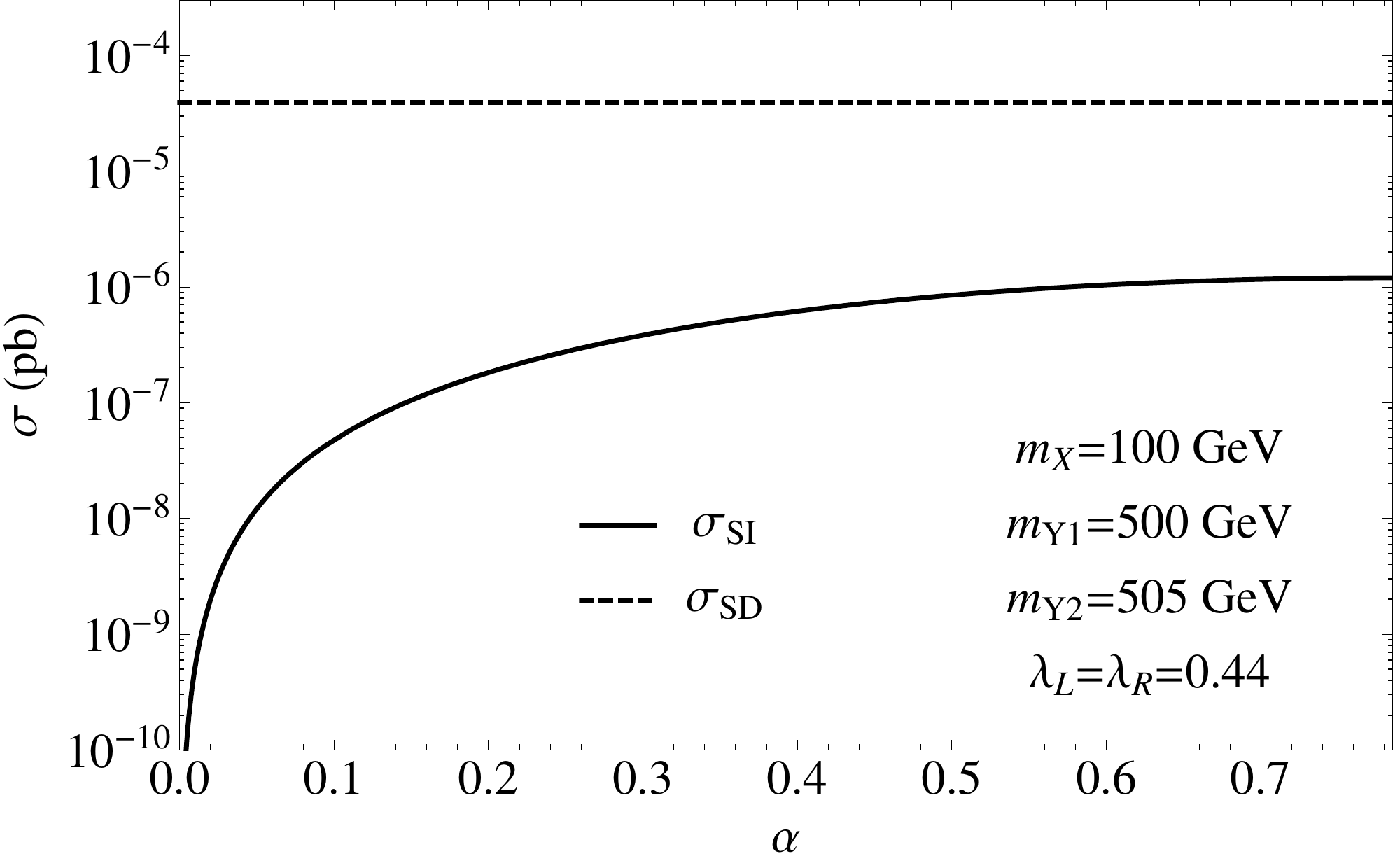}}\vspace{4mm}
\mbox{\includegraphics[width=0.48\textwidth]{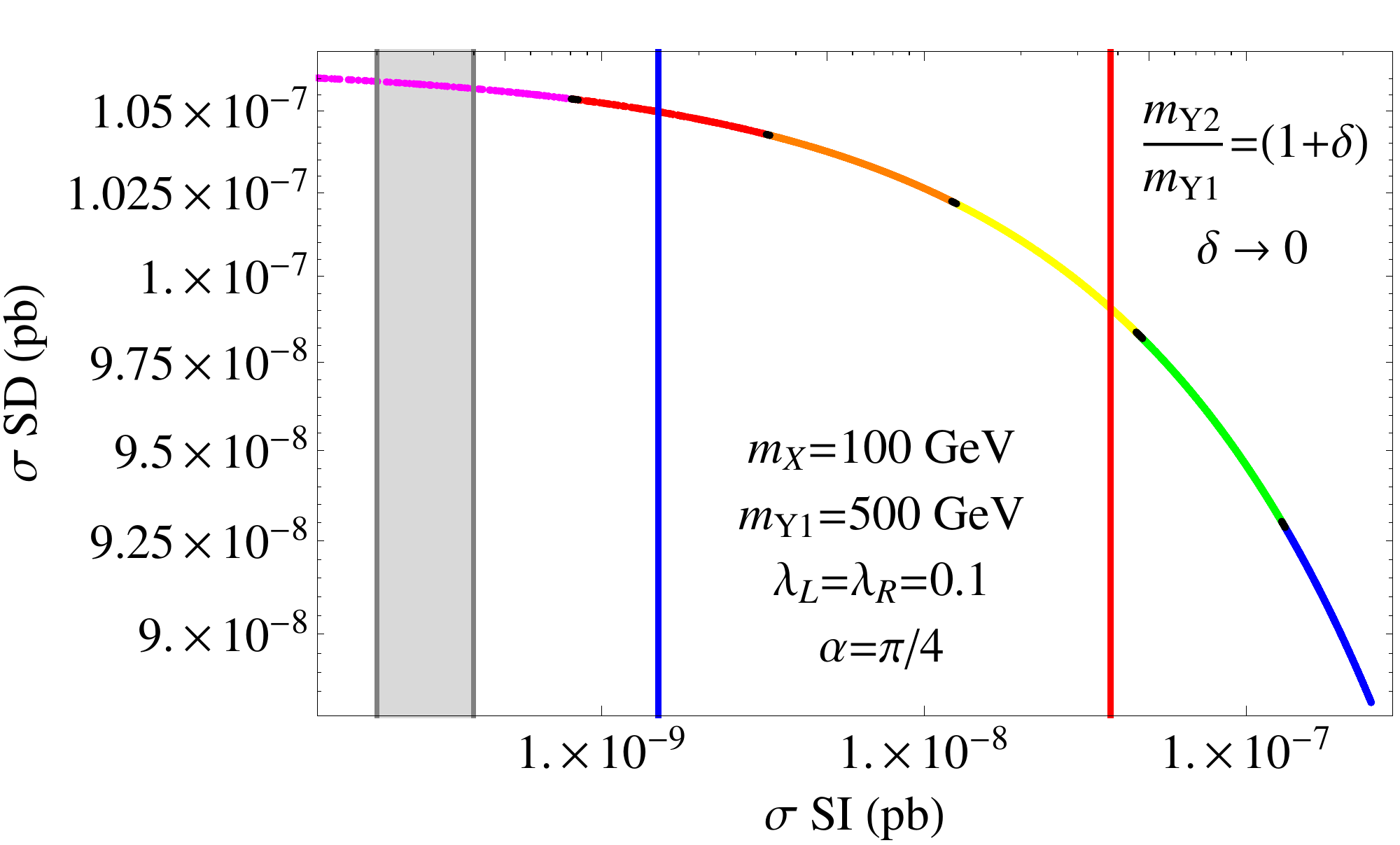}}\hspace{2mm}
\mbox{\includegraphics[width=0.48\textwidth]{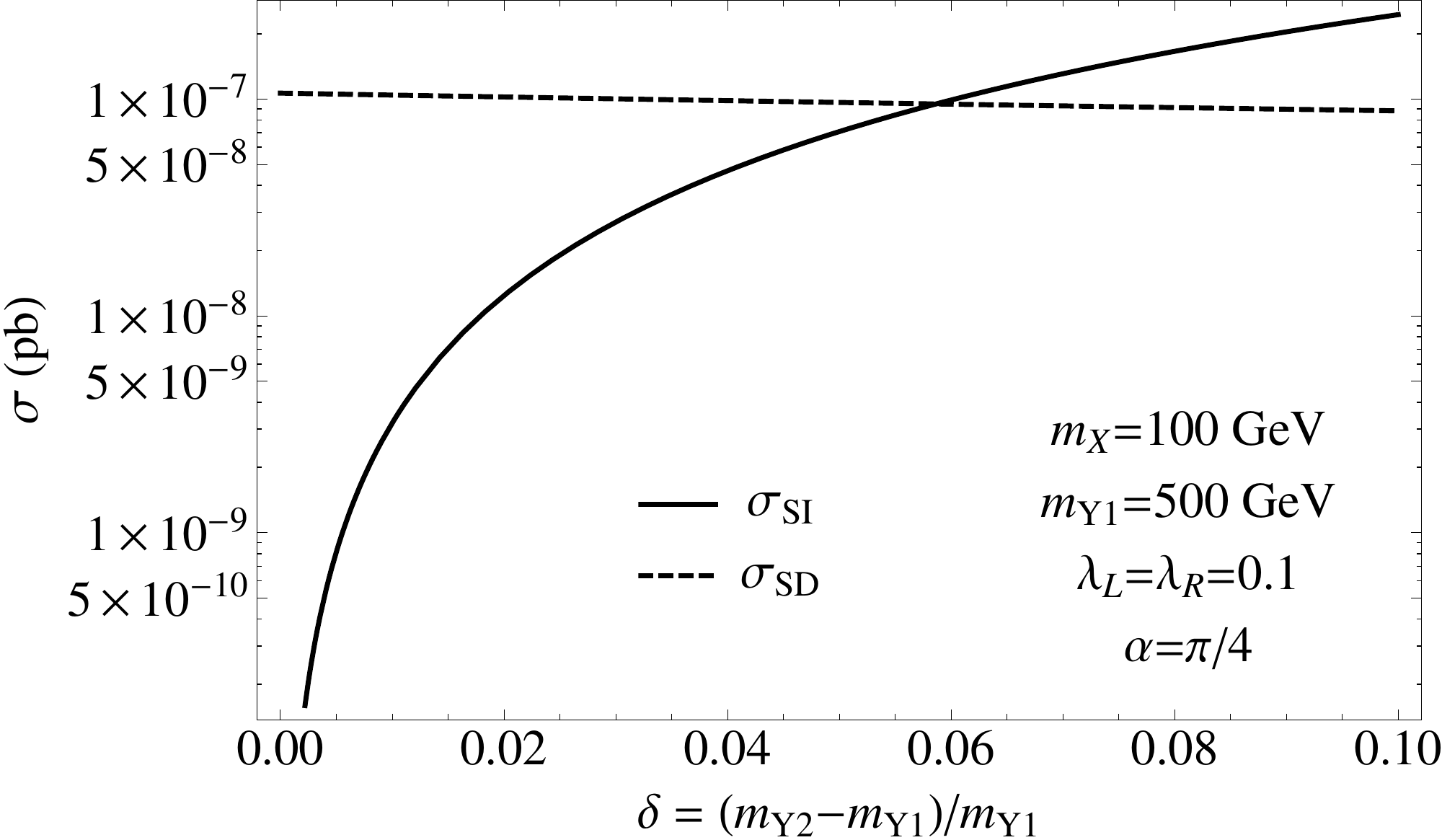}}
\end{center}
\caption{Departures from the limiting cases where $\sigmaSI \rightarrow 0$, as labeled in the upper
right corner of the left panels (vertical and horizontal lines denote the same experimental
sensitivities used in Fig.~\ref{fig:bigscan}).  In the left panels, we show the evolution in the $(\sigmaSI,\sigmaSD)$
plane as the small parameter ($\lambda_R$, $\alpha$, and $\delta$, from top to bottom) is increased. The color
coding indicates the value of the small parameter (increasing 
from left to right); 0 to 0.005 (magenta), 0.005 to 0.01 (red), 0.01 to 0.02 (orange),
0.02 to 0.04 (yellow), 0.04 to 0.07 (green), and 0.07 to 0.1 (blue). In the right panels,
we show $\sigmaSD$ (dashed curves)
and $\sigmaSI$ (solid curves) individually as functions of the small parameter.
\label{fig:limits}}
\end{figure}

In Fig.~\ref{fig:limits}, we present the behavior of $\sigmaSI$ and $\sigmaSD$ as we relax each of the
three cases in which $\sigmaSI \rightarrow 0$.
We can parameterize each of these limits by a small parameter, which goes to zero as $\sigmaSI \rightarrow 0$.  These
small parameters are $\lambda_R$, $\alpha$, and $\delta = (m_{\tilde Y_2}-m_{\tilde Y_1})/m_{\tilde Y_1}$.
As a benchmark, we consider $m_X = 100~\gev$, and
$m_{\tilde Y_1} = 500~\gev$ ($m_{\tilde Y_1} < m_{\tilde Y_2}$).

In the top panels of Fig.~\ref{fig:limits}, we examine the limit $\lambda_R \rightarrow 0$, for maximal mixing, $\delta=0.01$,
and $\lambda_L=0.44$. For these
parameter choices, $\lambda_R \lesssim 0.065$ pushes $\sigmaSI$ below the current bounds, while $\sigmaSD$ is still large
enough that the LHC might be sensitive to it. For $\lambda_L = 0.6$ however, and
all other parameters fixed, one obtains $\sigmaSD$ large enough to be detected by IceCube for $\lambda_R \gtrsim 0.08$, but all $\lambda_R \gtrsim 0.04$ are
already excluded by non-observation of spin-independent scattering. For a model to be visible at IceCube but to have evaded the current constraint on $\sigmaSI$,
$\lambda_L \gtrsim 0.605$, for these parameters, including maximal mixing.
If the mixing is not maximal and $\alpha < \pi/4$, viable models may have smaller $\lambda_L$ and larger $\lambda_R$.

From the top right
panel of Fig.~\ref{fig:limits}, as
from the shape of the contour in the left panel, it is clear that both $\sigmaSI$ and
$\sigmaSD$ increase with $\lambda_R$, though $\sigmaSI$ is more sensitive to changes in $\lambda_R$ for small $\lambda_R$.
Both $\sigmaSI$ and $\sigmaSD$ are
sensitive to $\lambda_L$ as well; $\sigmaSI$ is approximately proportional to $\lambda_L^2$, but
$\sigmaSD \propto \lambda_L^4$ for $\lambda_L \ll \lambda_R$ and is dominated by $\lambda_R$ for $\lambda_R \gg \lambda_L$.

Turning to the middle row of panels in Fig.~\ref{fig:limits}, we examine the dependence on the mixing angle, $\alpha$, for $\delta = 0.01$
and $\lambda_L = \lambda_R = 0.44$.
In both panels in the middle row it is clear that $\sigmaSD$ is independent of $\alpha$. From equation~\ref{eq:alphas}, we see that $\sigmaSD$
is independent of $\alpha$ for
$\lambda_L=\lambda_R$ and/or $m_{\tilde Y_1} = m_{\tilde Y_2}$, both of which are satisfied for the choice of parameters here. $\sigmaSI$,
by contrast, changes
rapidly at small mixing angles.  Again, increasing the Yukawa couplings increases both $\sigmaSD$ and $\sigmaSI$. For the particular
parameter choices here, the current bound on $\sigmaSI$ is evaded for $\alpha \lesssim 0.08$.

In the bottom row of panels in Fig.~\ref{fig:limits}, we assume maximal mixing ($\alpha = \pi/4$) and $\lambda_L = \lambda_R = 0.1$,
and examine the departure from degeneracy of the squark masses (assuming $\delta >0$). For this choice of
parameters, we see from
the bottom panel on the left that $\delta \lesssim \textrm{few}\times 10^{-2}$ is necessary to evade current bounds on
$\sigmaSI$; however $\sigmaSD$ is very small for Yukawa couplings in this range and will not be probed, even at the LHC.
In the right panel, one can
see that the primary effect of increasing $\delta$ is to increase $\sigmaSI$, while $\sigmaSD$ decreases slightly.
It is possible to have much larger $\sigmaSD$ and $\sigmaSI$ if the Yukawa couplings are much larger than $\lambda_L=\lambda_R=0.1$.
When the Yukawa couplings are increased, it is still possible to have $\sigmaSI$ below current bounds, but a much smaller
value of $\delta$ is required;
for $\lambda_L=\lambda_R=0.35,$ $\sigmaSD$ is within the IceCube sensitivity, but a squark mass non-degeneracy of
$\delta \lesssim \textrm{few}\times 10^{-3}$
is necessary so that $\sigmaSI$ is not already excluded.

\begin{figure}[ht]
\begin{center}
\mbox{\includegraphics[width=0.48\textwidth]{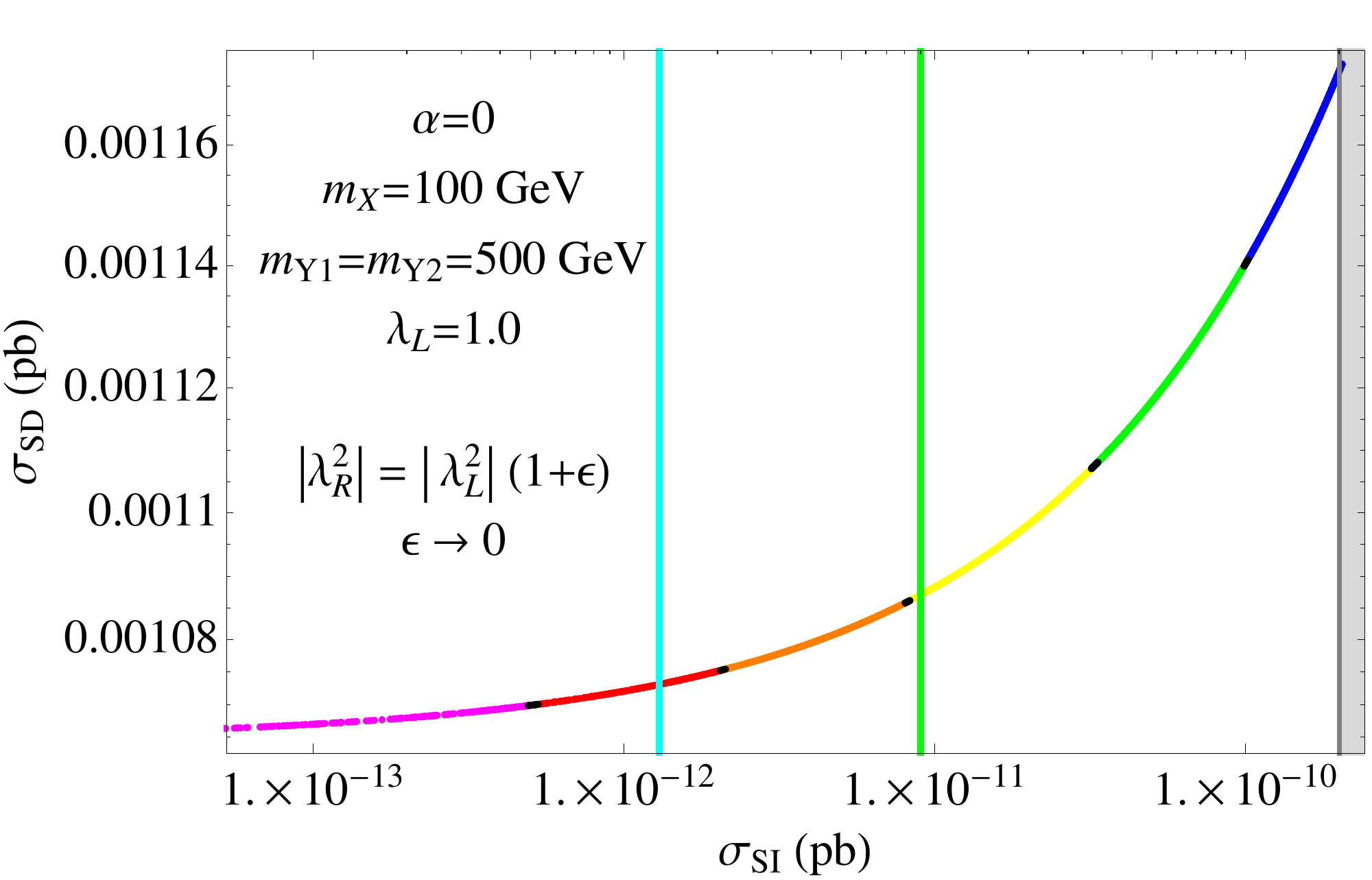}}\hspace{2mm}
\mbox{\includegraphics[width=0.48\textwidth]{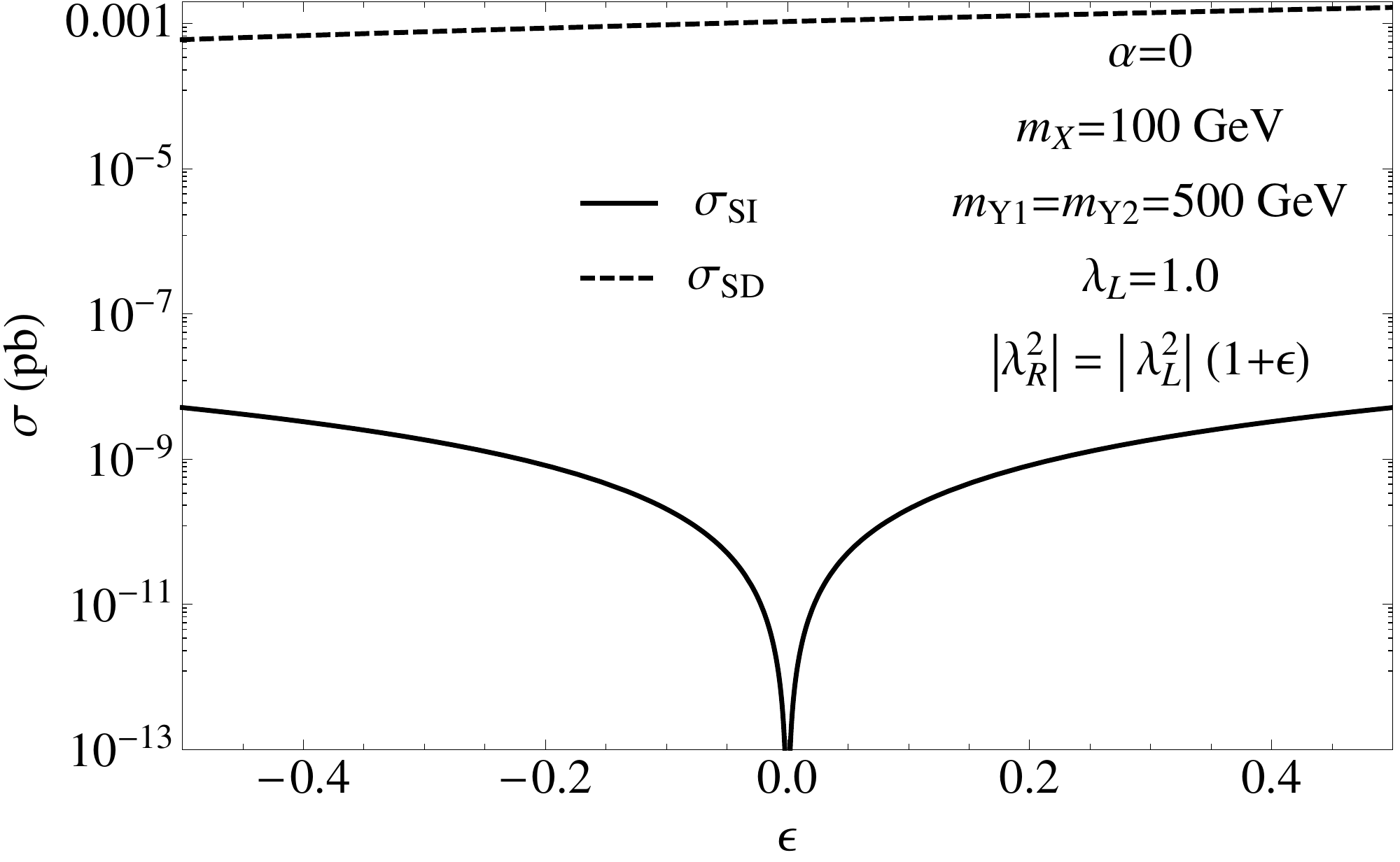}}
\end{center}
\caption{Departure from the limiting cases where $\sigmaSI \rightarrow 0$ due only to velocity-dependent terms (no squark mixing), as in Fig.~\ref{fig:limits}.
Again, vertical lines denote the experimental sensitivities used in Fig.~\ref{fig:bigscan}.
In the left panels, we show the evolution in the $(\sigmaSI,\sigmaSD)$
plane as $\epsilon$, the spilitting between $|\lambda_L^2|$ and $|\lambda_R^2|$, is increased.  The color
coding indicates the value of $\epsilon$ (increasing from left to right); 0 to 0.005 (magenta), 0.005 to 0.01 (red), 0.01 to 0.02 (orange),
0.02 to 0.04 (yellow), 0.04 to 0.07 (green), and 0.07 to 0.1 (blue). In the right panel,
we show $\sigmaSD$ (dashed curves)
and $\sigmaSI$ (solid curves) individually as functions of $\epsilon$.
\label{fig:velocityonly}}
\end{figure}

In Figure~\ref{fig:velocityonly}, we examine the possibility that there is no squark mixing ($\alpha = 0$).  In this case, the spin-independent
elastic scattering cross section is due entirely to the velocity-suppressed contributions $\sigmaSI^{(1)}$ and $\sigmaSI^{(6)}$.
Since we have assumed real Yukawa couplings, $\sigmaSI^{(6)}=0$.
We see from the figure that even for $\lambda_L \approx \lambda_R \approx 1$, $\sigmaSI$ is well below the current constraints, while $\sigmaSD$ is quite large\footnote{Note that $\sigmaSD$ is only mildly dependent on $\epsilon$, as evident in the right panel of Fig.~\ref{fig:velocityonly}. For the parameter choices in the left panel, none of the $\sigmaSD$ experimental sensitivities are visible, though the range of $\sigmaSD$ plotted here lies just below the Tevatron sensitivity (shown as a horizontal green dotted line in Fig.~\ref{fig:bigscan}) 
and well-above the sensitivity of Large DMTPC and IceCube/DeepCore (shown as grey solid and cyan dashed lines, respectively, in Fig.~\ref{fig:bigscan}).}.
For $\epsilon > 0$, both $\sigmaSD$ and $\sigmaSI$ increase with $\epsilon$, though only $\sigmaSI \rightarrow 0$ rapidly as $\epsilon \rightarrow 0$.
Here, $\epsilon \lesssim 0.008$ will result in dark matter that will be evident only through its spin-dependent scattering.
All other parameters fixed, models can be visible at both IceCube and LUX/ZEP 20T with $\lambda_L$ as small as 0.421, with $\epsilon >0.045$ ($\lambda_R > 0.412$).


Finally, returning to the parameter scan in Fig.~\ref{fig:bigscan}, of great interest are the regions of parameter space
in which a discovery may be made by IceCube/DeepCore or the LHC,
but which would evade all searches for spin-independent scattering on nuclei, namely those points with
$\sigmaSD \gtrsim 10^{-6}$ pb and $\sigmaSI \lesssim 10^{-12}$ pb.
These points are characterized by large $m_{\tilde Y_1}$ and $m_{\tilde Y_2}$ ($\gtrsim 1500$ GeV, near
the edge of our range), one large Yukawa coupling ($\gtrsim 2$),
and one small Yukawa coupling ($\lesssim 0.2$), {\it or} very nearly degenerate squark masses ($\delta \approx 0$)
and two relatively large Yukawa couplings, {\it or} a
combination of near-degeneracy of squark masses and large-ish Yukawa couplings.  In the latter cases,
small-ish $\alpha$ can also help induce a small $\sigmaSI$ without
reducing $\sigmaSD$ due to the degeneracy of the squark masses. In the first case, if the small Yukawa is
very small ($\lesssim 10^{-2}$), the the lighter squark can be
as light as $\sim1$ TeV, and the larger Yukawa does not necessarily have to be larger than 2.

This result is not unexpected; in the limit of small $\alpha$ and $\delta$,
${\alpha_3 \over \alpha_2} \propto (\lambda_R / \lambda_L )^2 \alpha \delta $,
so a scan with linear priors will favor the region of parameter-space with large
Yukawa couplings and masses.  Logarithmic priors would instead favor all mass and coupling scales equally.
Given linear priors and the suppression needed to evade upcoming bounds on $\sigmaSI$, the points we have
generated which may be
detected by spin-dependent scattering but which will evade
detection by spin-independent scattering occur when at least two
of the $\sigmaSI \rightarrow 0$ criteria are approximately satisfied.

In this analysis, we have restricted the WIMPless dark matter mass to be $m_X=100$ GeV due to the wealth of direct dark matter searches sensitive to this mass,
however a similar analysis may be carried out for any $m_X$.  Both $\sigmaSD$ and $\sigmaSI$ decrease for smaller $m_X$, but the changes to Fig.~\ref{fig:bigscan}
are minor, with the exception of the relevant constraints. Indeed, the qualitative conclusions drawn from Figs.~\ref{fig:limits} and~\ref{fig:velocityonly} are valid for any $m_X$.

\section{Conclusion}

We have considered the detection prospects for Majorana fermion WIMPless dark matter at current and near future
dark matter detectors, using both spin-independent and spin-dependent scattering.  We have found that although Majorana fermion WIMPless dark
matter always exhibits spin-dependent nuclear scattering, spin-independent scattering contributions are always either velocity-dependent or
dependent on mixing of exotic 4th generation squarks.  One-loop corrections do not generate spin-independent contributions
which could potentially be detected at upcoming experiments.

A scan of models shows that the majority of models which could be detected by IceCube/DeepCore (with 1800 live-days of
data) or at the LHC can also be detected through spin-independent scattering by detectors such as SuperCDMS, XENON100 and DEAP/CLEAN (as well as the first
generation LUX detector) over a
comparable time frame.  However, a significant fraction of models detectable at IceCube/DeepCore or the LHC would not be detected
through spin-independent scattering until major future upgrades are made, if at all.  Unlike the case of MSSM neutralino WIMPs, it is
possible for WIMPless Majorana fermions which have evaded detection even through velocity-suppressed spin-independent scattering to
nevertheless be detected by current detectors through spin-dependent scattering.

\acknowledgements
We gratefully acknowledge J.~L.~Feng for collaboration at an early stage of this project, and
M.~Felizardo, D.~Marfatia and T.~Tait for useful discussions.  This work is supported in part by DOE grant DE-FG02-04ER41291.
P.S.~is supported by the National Science Foundation under Grant Numbers PHY-0969020 and PHY-0455649 and
would also like to thank the University of Hawai'i for hospitality during the course of this research.



\end{document}
